\documentclass[useAMS,usenatbib,onecolumn]{mn2e}
\usepackage{amsmath,bm}
\usepackage{dcolumn}
\usepackage[utf8]{inputenc}
\usepackage{graphics,graphicx}
\usepackage{float}
\usepackage{threeparttable}
\usepackage{url}
\usepackage{epstopdf}

\usepackage{color}

\newcommand{\p}{^\prime}
\newcommand{\pp}{^{\prime\prime}}

\newcommand{\ket}[1]{|#1\rangle}

\title[ExoMol line lists -- XXIX. CH$_3$Cl]{ExoMol line lists -- XXIX. The rotation-vibration spectrum of methyl chloride up to 1200\,K}

\date{\today}

\author[A. Owens et al.]
{A. Owens,$^{1,2}$\thanks{The corresponding author: alec.owens@cfel.de} A. Yachmenev,$^{1,2}$ W. Thiel,$^3$ A. Fateev,$^{4}$ \newauthor J. Tennyson$^5$\thanks{The corresponding author: j.tennyson@ucl.ac.uk} and S. N. Yurchenko$^5$\thanks{The corresponding author: s.yurchenko@ucl.ac.uk}\vspace*{4mm}\\
$^1$ The Hamburg Center for Ultrafast Imaging, Universit\"{a}t Hamburg, Luruper Chaussee 149, 22761 Hamburg, Germany\\
$^2$ Center for Free-Electron Laser Science (CFEL), Deutsches Elektronen-Synchrotron DESY, Notkestrasse 85, 22607 Hamburg, Germany\\
$^3$ Max-Planck-Institut f\"{u}r Kohlenforschung, Kaiser-Wilhelm-Platz 1, 45470 M\"{u}lheim an der Ruhr, Germany\\
$^4$ Technical University of Denmark, Department of Chemical and Biochemical Engineering, Frederiksborgvej 399, 4000 Roskilde, Denmark\\
$^5$ Department of Physics and Astronomy, University College London, Gower Street, WC1E 6BT London, United Kingdom}

\date{Accepted XXXX. Received XXXX; in original form XXXX}

\pagerange{\pageref{firstpage}--\pageref{lastpage}} \pubyear{2018}

\begin{document}

\label{firstpage}

\maketitle

\begin{abstract}
Comprehensive rotation-vibration line lists are presented for the two main isotopologues of methyl chloride, $^{12}$CH$_3{}^{35}$Cl and $^{12}$CH$_3{}^{37}$Cl. The line lists, OYT-35 and OYT-37, are suitable for temperatures up to $T=1200\,$K and consider transitions with rotational excitation up to $J=85$ in the wavenumber range $0$\,--\,$6400\,$cm$^{-1}$ (wavelengths $\lambda> 1.56\,\mu$m). Over 166 billion transitions between 10.2 million energy levels have been calculated variationally for each line list using a new empirically refined potential energy surface, determined by refining to 739 experimentally derived energy levels up to $J=5$, and an established \textit{ab initio} dipole moment surface. The OYT line lists show excellent agreement with newly measured high-temperature infrared absorption cross-sections, reproducing both strong and weak intensity features across the spectrum. The line lists are available from the ExoMol database and the CDS database.
\end{abstract}


\section{Introduction}

 The recent interstellar detection of methyl chloride around the protostar IRAS 16293-2422 and in the coma of comet 67P/Churyumov-Gerasimenko (67P/C-G)~\citep{17FaObJo.CH3Cl} has undermined the possibility of CH$_3$Cl as a realistic biosignature gas in the search for life outside of our Solar system~\citep{05SeKaMe.CH3Cl,13SeBaHu.CH3Cl,13aSeBaHu.CH3Cl}. The fact that CH$_3$Cl can be formed abiotically in these environments, and possibly delivered by cometary impact to young planets, means it is now far more relevant in the context of newly formed rocky exoplanets. Consequently, there is renewed incentive for a comprehensive rotation-vibration line list of methyl chloride that is suitable for elevated temperatures.

 Since 2012, the ExoMol database~\citep{ExoMol2012,ExoMol2016} has been generating molecular line lists and key spectroscopic data on a variety of small molecules deemed important for the characterization of hot astronomical atmospheres. Notable applications utilizing ExoMol line lists include: the use of the 10to10 line list~\citep{14YuTexx.CH4} to model methane in exoplanets~\citep{jt495,14YuTeBa.CH4,jt699} and the bright T4.5 brown dwarf 2MASS 0559-14~\citep{14YuTeBa.CH4}, and to assign lines in the near-infrared spectra of late T dwarfs~\citep{jt596} in combination with the ammonia BYTe line list~\citep{jt500}; the early detection of water using the BT2 line list~\citep{jt378} in HD 189733b~\citep{jt400} and HD 209458b~\citep{jt488}; and the provisional identification of HCN in the atmosphere of super-Earth 55 Cancri e~\citep{jt629} and TiO in the atmosphere of hot Jupiter WASP-76 b~\citep{jt699}. Conversely, the possible detection of NaH in the atmosphere of a brown dwarf was ruled out using a line list for this molecule~\citep{jt605}.

 In this work, we present newly computed rotation-vibration line lists, named OYT-35 and OYT-37, for the two main isotopologues of methyl chloride, $^{12}$CH$_3{}^{35}$Cl and $^{12}$CH$_3{}^{37}$Cl (henceforth referred to as CH$_3{}^{35}$Cl and CH$_3{}^{37}$Cl). These line lists are validated against new high-temperature infrared (IR) absorption cross-sections measured at temperatures up to 500$\,^{\circ}{\rm C}$. Methyl chloride is the fourth pentatomic molecule to be considered within the ExoMol framework~\citep{ExoSoft2016} after the 10to10 line list of CH$_4$~\citep{14YuTexx.CH4}, the line list of HNO$_3$~\citep{Pavlyuchko2015} and the OY2T line list of SiH$_4$~\citep{17OwYaTh.SiH4}. The OYT line lists are a continuation of our previous efforts where we constructed potential energy and dipole moment surfaces for CH$_3$Cl using state-of-the-art electronic structure theory~\citep{15OwYuYa.CH3Cl,16OwYuYa.CH3Cl}. Variational nuclear motion calculations were used to rigorously evaluate these surfaces and they were shown to display excellent agreement with a range of experimental spectroscopic data. Notably, band shape and structure was well reproduced across the IR spectrum of CH$_3$Cl, even for weaker intensity features.

 The paper is structured as follows: In Sec.~\ref{sec:methods}, the theoretical approach and experimental setup are presented. This includes details on the empirical refinement of the potential energy surface (PES), the dipole moment surface (DMS) and intensity simulations, and the variational nuclear motion calculations. The OYT line lists are described in Sec.~\ref{sec:results}, where we look at the temperature-dependent partition functions of CH$_3{}^{35}$Cl and CH$_3{}^{37}$Cl, the format and temperature dependence of the OYT line lists, and comparisons with the HITRAN database and IR absorption cross-sections measured at temperatures up to 500$\,^{\circ}{\rm C}$. We conclude in Sec.~\ref{sec:conc}.

\section{Methods}
\label{sec:methods}

\subsection{Potential energy surface refinement}

 The CBS-35$^{\,\mathrm{HL}}$ PES~\citep{15OwYuYa.CH3Cl} utilized in this work is based on extensive, high-level \textit{ab initio} calculations. Despite reproducing the fundamentals with a root-mean-square (rms) error of $0.75\,$cm$^{-1}$, orders-of-magnitude improvements in the accuracy of the predicted transition frequencies can be obtained through empirical refinement of the PES. Improved energy levels also results in better wavefunctions and more reliable intensities. Since refinement is computationally intensive, the PES was only refined to CH$_3{}^{35}$Cl experimental term values, which is the main isotopologue. As we will see in Sec.~\ref{sec:results}, the resultant refined PES is still suitable for CH$_3{}^{37}$Cl.

 The refinement was performed using an efficient least-squares fitting procedure~\citep{YuBaTe11.NH3} implemented in the nuclear motion program \textsc{trove}~\citep{TROVE2007}. %Assuming the CBS-35$^{\,\mathrm{HL}}$ PES, $V_{\rm CBS\mbox{-}35^{HL}}$, is a reasonable starting point, the effect of the refinement can be treated as a perturbation $\Delta V$ such that
To make the procedure more computationally tractable for CH$_3$Cl, the number of expansion parameters of the CBS-35$^{\,\mathrm{HL}}$ PES was carefully reduced from 414 to 188 without significant loss in accuracy. Of the 188 parameters, only 33 were eventually varied in the refinement. A total of 739 experimental term values up to $J=5$ were used and this included 52 vibrational $J=0$ band centres taken from \citet{HITRAN2012,05NiChBu.CH3Cl,90DuLaxx.CH3Cl,99Laxxxx.CH3Cl,11BrPeJa.CH3Cl}. All $J>0$ energies were taken from the HITRAN2012 database~\citep{HITRAN2012} apart from the pure rotational energies of \citet{05NiChxx.CH3Cl}. Experimental energy levels from HITRAN are fairly robust as they are usually derived from numerous observed transitions, unlike for example, so-called dark states which are harder to determine. In particular, we had difficulty refining to the $2\nu_5(E)$, $2\nu_3+\nu_5(E)$ and $3\nu_6(A_1)$ $J=0$ states between $\approx2895$\,--\,$3060\,$cm$^{-1}$ from \citet{11BrPeJa.CH3Cl} and their weights in the refinement were subsequently reduced to lessen their influence on the final PES. Pure rotational energy levels had the largest weights, whilst term values from HITRAN were weighted two orders-of-magnitude smaller. The remaining $J=0$ energy levels not present in HITRAN, i.e. those from \citet{90DuLaxx.CH3Cl,99Laxxxx.CH3Cl}, were given an order-of-magnitude smaller weights. Relative weighting is more important in the refinement than the absolute values since the weights are normalized (see \citet{YuBaTe11.NH3} for further details).

 Figure~1 displays the results of the refinement through the fitting residuals $\Delta E({\rm obs-calc})=E_{\rm obs}-E_{\rm calc}$, where $E_{\rm obs}$ and $E_{\rm calc}$ are the observed and calculated energies, respectively. The rms error for the 706 term values from HITRAN2012 is $0.072\,$cm$^{-1}$ and this level of accuracy extends across all $J$. The remaining 33 $J=0$ wavenumbers possess much larger residual errors and although this is partly due to the weighting, in certain cases we would have expected better agreement at lower energies. We therefore believe that the accuracy of some of the experimentally determined $J=0$ levels can be improved. Unfortunately, we were unable to incorporate new data from the latest HITRAN2016~\citep{HITRAN2016} release, predominantly involving analysis of the $1900$\,--\,$2600\,$cm$^{-1}$ spectral region~\citep{16NiDmGo.CH3Cl}, because the refinement was completed before this became available to us.

\begin{figure}
\centering
\label{fig:refine}
\includegraphics[width=0.47\textwidth]{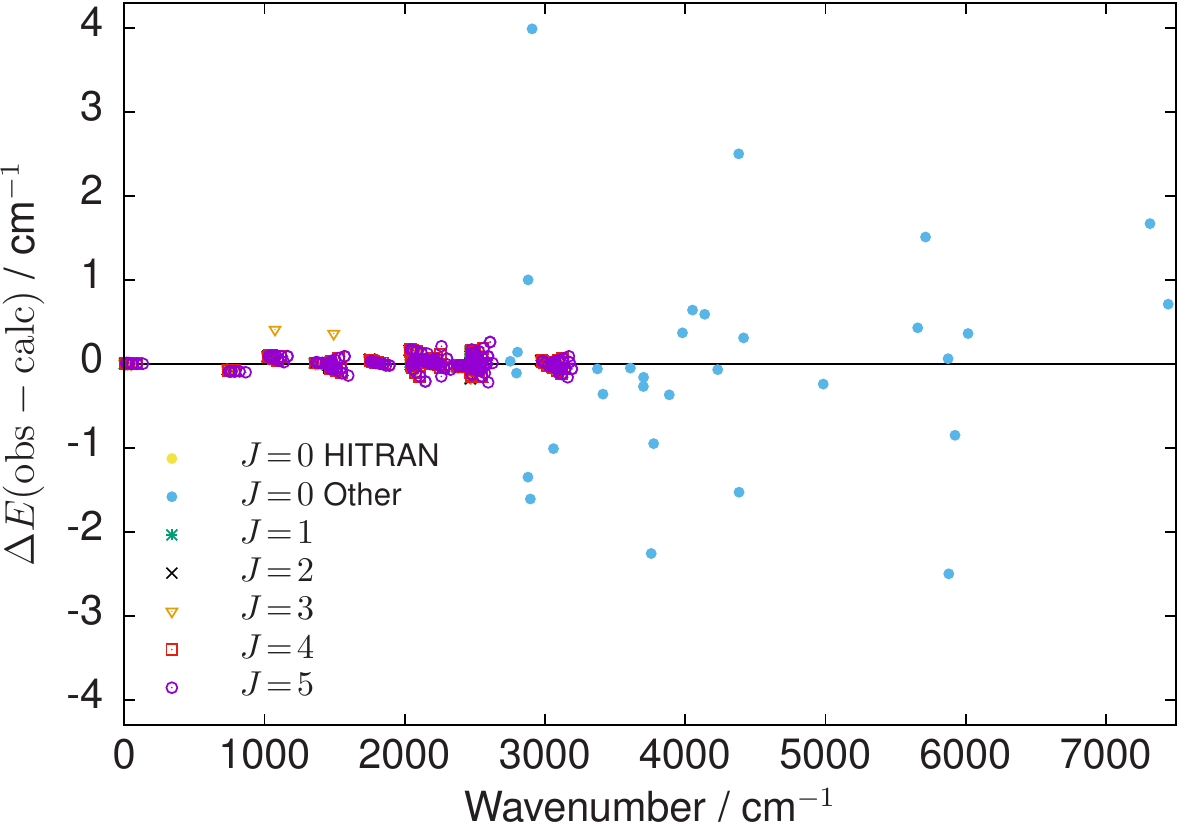}
\caption{Fitting residuals, $\Delta E({\rm obs-calc})=E_{\rm obs}-E_{\rm calc}$, of the 739 energy levels used in the PES refinement.}
\end{figure}

 It should be noted that the accuracy of the refined PES can only be guaranteed with the exact  computational setup used in this study. This is to be expected in theoretical line list production using programs that do not treat the kinetic energy operator exactly, e.g. \textsc{trove}. The refined PES is not recommended for future use but is provided in the supplementary material along with a Fortran routine to construct it.

\subsection{Dipole moment surface and line intensities}

 The electric DMS used in this work was generated at the CCSD(T)/aug-cc-pVQZ(+d for Cl) level of theory and the reader is referred to \citet{16OwYuYa.CH3Cl} for a detailed description and evaluation of this surface. Absolute absorption intensities were simulated using the expression,
\begin{equation}
\label{eq:abs_I}
I(f \leftarrow i) = \frac{A_{if}}{8\pi c}g_{\mathrm{ns}}(2 J_{f}+1)\frac{\exp\left(-E_{i}/kT\right)}{Q(T)\; \nu_{if}^{2}}\left[1-\exp\left(-\frac{hc\nu_{if}}{kT}\right)\right] ,
\end{equation}
where $A_{if}$ is the Einstein-$A$ coefficient of a transition with wavenumber $\nu_{if}$ (in cm$^{-1}$) between an initial state with energy $E_i$ and a final state with rotational quantum number $J_f$. Here, $k$ is the Boltzmann constant, $h$ is the Planck constant, $c$ is the speed of light and $T$ is the absolute temperature. The nuclear spin statistical weights of both isotopologues are $g_{\mathrm{ns}}=\lbrace 16,16,16\rbrace$ for states of symmetry $\lbrace A_1,A_2,E\rbrace$, respectively, and $Q(T)$ is the temperature-dependent partition function. Transitions follow the symmetry selection rules $A_1 \leftrightarrow A_2,\; E \leftrightarrow E$; and the standard rotational selection rules, $J\p-J\pp=0,\pm 1,\; J\p+J\pp \ne 0$; where $\p$ and $\pp$ denote the upper and lower state, respectively. All spectral simulations were carried out with the \textsc{ExoCross} code~\citep{ExoCross:2018}.

\subsection{Variational calculations}

 Variational calculations were performed with \textsc{trove}, whose methodology has been well documented~\citep{TROVE2007,09YuBaYa.NH3,15YaYu.ADF,ExoSoft2016,Symmetry:2017}. Since rovibrational computations on CH$_3$Cl have previously been reported~\citep{15OwYuYa.CH3Cl,16OwYuYa.CH3Cl}, we describe only the key details relevant for this work.

 An automatic differentiation method~\citep{15YaYu.ADF} was used to construct the rovibrational Hamiltonian, which was represented as a power series expansion around the equilibrium geometry in terms of nine, curvilinear internal coordinates. The kinetic and potential energy operators  were both truncated at sixth order. Atomic mass values were used throughout. The symmetrized vibrational basis set was generated using a multi-step contraction scheme~\citep{Symmetry:2017} and the size was controlled by the polyad number
\begin{equation}\label{eq:polyad_sih4}
P = 2(n_1+n_2+n_3+n_4)+n_5+n_6+n_7+n_8+n_9 \leq P_{\mathrm{max}} .
\end{equation}
The quantum numbers $n_k$ for $k=1,\ldots,9$ correspond to the primitive basis functions $\phi_{n_k}$, which are determined by solving one-dimensional Schr\"{o}dinger equations for each $k$th vibrational mode using the Numerov-Cooley method~\citep{Numerov1924,Cooley1961}. Multiplication with symmetrized rigid-rotor eigenfunctions $\ket{J,K,m,\tau_{\mathrm{rot}}}$ gives the final basis set for $J>0$ calculations. The quantum numbers $K$ and $m$ are the projections (in units of $\hbar$) of $\hat{J}$ onto the molecule-fixed $z$ axis and the laboratory-fixed $Z$ axis, respectively, whilst $\tau_{\mathrm{rot}}$ determines the rotational parity as $(-1)^{\tau_{\mathrm{rot}}}$.

 Initially, $J=0$ calculations were done with $P_{\mathrm{max}}=12$, which resulted in 49\,076 vibrational basis functions with energies up to $h c \cdot 20$\,600\,cm$^{-1}$. With such a large basis set, describing rotational excitation quickly becomes computationally intractable and it was therefore necessary to reduce the number of basis functions. This was done using a basis set truncation procedure based on the vibrational transition moments. 
%, which are defined as
%\begin{equation}
%\label{eq:TM}
%\mu_{if} = \sqrt{\sum_{\alpha=x,y,z}{\lvert\bra{\Phi^{(f)}_{\mathrm{vib}}}\bar{\mu}_{\alpha}\ket{\Phi^{(i)}_{\mathrm{vib}}}\rvert}^2} ,
%\end{equation}
%where $\ket{\Phi^{(i)}_{\mathrm{vib}}}$ and $\ket{\Phi^{(f)}_{\mathrm{vib}}}$ are the initial and final state vibrational eigenfunctions, respectively, and $\bar{\mu}_{\alpha}$ is the electronically averaged dipole moment function along the molecule-fixed axis $\alpha=x,y,z$.
These are relatively inexpensive to compute in \textsc{trove} and such a scheme was previously employed when generating a comprehensive line list for SiH$_4$~\citep{17OwYaTh.SiH4}. All possible transition moments were calculated for a lower state energy threshold of $h c \cdot 8000\,$cm$^{-1}$ (same as for the OYT line list intensity calculations), from which vibrational band intensities were estimated at an elevated temperature, e.g. $1500\,$K. For each $J=0$ energy level, and thus each basis function, a band intensity value was assigned which was simply the largest value computed for that state. The vibrational basis set was then reduced by removing basis functions, and energy levels, above $h c \cdot 8000\,$cm$^{-1}$ with band intensity values smaller than $3\times 10^{-22}\,$cm/molecule. This is around three orders of magnitude smaller than the largest computed value and was chosen primarily for computational reasons. The final pruned basis sets contained 2158 and 2156 vibrational basis functions with energies up to $h c \cdot 10$,400\,cm$^{-1}$ for CH$_3{}^{35}$Cl and CH$_3{}^{37}$Cl, respectively. They were multiplied in the usual manner with symmetrized rigid-rotor functions for $J>0$ calculations. Naturally, by using this truncation procedure we will lose information on weaker lines involving states above $h c \cdot 8000\,$cm$^{-1}$ and without more rigorous calculations the exact effects are hard to quantify. Predicted rovibrational energies are also affected, however, we have compensated for this error to some extent by refining the PES with the pruned basis set.

 The OYT line lists were computed with a lower state energy threshold of $h c \cdot 8000\,$cm$^{-1}$ and considered transitions up to $J=85$ in the $0$\,--\,$6400\,$cm$^{-1}$ range. The accuracy of the line lists was further improved by performing an empirical basis set correction~\citep{09YuBaYa.NH3}, which involves a shift of the vibrational band centres to better match experiment. For both isotopologues of CH$_3$Cl, we replaced sixteen band centres up to $2500\,$cm$^{-1}$ with values from \citet{05NiChBu.CH3Cl} and three band centres around the $3000\,$cm$^{-1}$ region with values from \citet{11BrPeJa.CH3Cl} (see Table~\ref{tab:ebsc}).
\begin{table}
\centering
\tabcolsep=11pt
\caption{\label{tab:ebsc}Observed vibrational band centres $\nu$ (in cm$^{-1}$) from \citet{05NiChBu.CH3Cl} and \citet{11BrPeJa.CH3Cl} used in the empirical basis set correction of the OYT line lists.}
\begin{tabular}{l c c c}
\hline\hline\\[-3.0mm]
Mode & Sym. & $\nu$ (CH$_3{}^{35}$Cl) & $\nu$ (CH$_3{}^{37}$Cl) \\[0.5mm]
\hline\\[-3.0mm]
$\nu_3$       & $A_1$  & \,\,\,732.84& \,\,\,727.03\\
$\nu_6$       & $E$    &   1018.07&  1017.68\\
$\nu_2$       & $A_1$  &   1354.88&  1354.69\\
$\nu_5$       & $E$    &   1452.18&  1452.16\\
$2\nu_3$      & $A_1$  &   1456.76&  1445.35\\
$\nu_3+\nu_6$ & $E$    &   1745.37&  1739.24\\
$2\nu_6$      & $A_1$  &   2029.38&  2028.59\\
$2\nu_6$      & $E$    &   2038.33&  2037.56\\
$\nu_2+\nu_3$ & $A_1$  &   2080.54&  2074.45\\
$3\nu_3$      & $A_1$  &   2171.89&  2155.12\\
$\nu_3+\nu_5$ & $E$    &   2182.57&  2176.75\\
$\nu_2+\nu_6$ & $E$    &   2367.72&  2367.14\\
$\nu_5+\nu_6$ & $E$    &   2461.65&  2461.48\\
$2\nu_3+\nu_6$& $E$    &   2463.82&  2451.90\\
$\nu_5+\nu_6$ & $A_1$  &   2464.90&  2464.47\\
$\nu_5+\nu_6$ & $A_2$  &   2467.67&  2467.25\\
$\nu_1$       & $A_1$  &   2967.77&  2967.75\\
$\nu_4$       & $E$    &   3037.14&  3036.75\\
$3\nu_6$      & $E$    &   3045.02&  3044.14\\[0.5mm]
\hline\hline
\end{tabular}
\end{table}

\subsection{Experimental setup}
\label{sec:experiment}

 Mid IR CH$_3$Cl absorption measurements were carried out in the $600$\,--\,$6000\,$cm$^{-1}$ region for temperatures between 22\,--\,500$\,^{\circ}{\rm C}$ ($295.15$\,--\,$773.15\,$K) at a pressure of about 1 bar using a quartz flow gas cell as previously described in \citet{Grosch2013}. Full details on the optical set up and raw data analysis can be found in \citet{Barton2015}. Measurements used an Agilent 660 spectrometer with $0.09$\,--\,$2\,$cm$^{-1}$ spectral resolution and a linearized broad-band mercury cadmium telluride (MCT) detector. The spectra were calculated from the measured interferograms using triangular apodization functions. The Lambert-Beer law was used for all absorption spectral calculations.

 Pressurized CH$_3$Cl (99.8\%) from Air Liquide was diluted with N$_2$ (99.998\%) to obtain a CH$_3$Cl concentration in N$_2$ at the few vol\,\% level. The CH$_3$Cl and N$_2$ flows were controlled with high-end mass-flow controllers (MFC). The actual CH$_3$Cl concentrations have been calculated based on known CH$_3$Cl and N$_2$ flows (defined by the MFC) and were in the $1$\,--\,$7$ vol\,\% range. As a check of our experimental setup, measured spectra at 25$\,^{\circ}{\rm C}$ (resolution of $0.09\,$cm$^{-1}$) were compared with experimental data from the PNNL database~\citep{PNNL}, which employs a similar class Fourier Transform Infrared (FTIR) spectrometer in their measurements. The obtained results showed very good agreement with the PNNL database. Because CH$_3$Cl absorption features are relatively broad compared to a resolution of $0.09\,$cm$^{-1}$, most experiments were carried out with $0.5\,$cm$^{-1}$ resolution to achieve better signal-to-noise (S/N) ratio and measurement time. Experimental cross-sections of CH$_3$Cl in the region $600$\,--\,$6000\,$cm$^{-1}$ are provided as supplementary material. This includes measurements for temperatures of 25, 300, 400 and 500$\,^{\circ}{\rm C}$ at a pressure of about 1 bar and resolution of $0.5\,$cm$^{-1}$.

 The CH$_3$Cl IR absorption cross-sections are relatively low in the mid IR ($\approx 2.5 \times 10^{-20}\,$cm$^2$/molecule). In practical applications these measurements are quite complicated because of possible spectral interferences with H$_2$O, CO$_2$ and H$_x$C$_y$. In general, relatively long (few metres) absorption path lengths are required for CH$_3$Cl measurements at the ppm-level but these difficulties can be overcome, for example, by employing a different spectral range. Molecules often possess relatively large absorption cross-sections below $200\,$nm, meaning the path length can be significantly reduced when measuring far ultraviolet (UV) or vacuum ultraviolet (VUV) spectra.

\section{Results}
\label{sec:results}

\subsection{OYT line list format}

 The ExoMol data structure has been adopted for the OYT line lists and a detailed description with illustrative examples can be found in \citet{ExoMol2016}. The \texttt{.states} file, see Table~2, includes all the computed rovibrational energies (in cm$^{-1}$), with each energy level possessing a unique state counting number, symmetry and quantum number labelling, and the contribution $|C_i|^2$ from the largest eigen-coefficient used to assign the rovibrational state. The \texttt{.trans} files are split into $100\,$cm$^{-1}$ frequency bins for handling purposes and contain all computed transitions with upper and lower state ID labels and Einstein-$A$ coefficients, as shown in Table~3. 
 
 The \textsc{trove} (local mode) vibrational quantum numbers $n_1,\ldots,n_9$ can be related to the normal mode quantum numbers ${\rm v}_1,\ldots{\rm v}_6$ as follows:
\begin{eqnarray*}
  {\rm v}_3  &=& n_1,  \quad {\rm CCl-stretch},   \\
  {\rm v}_1+{\rm v}_4  &=& n_2 + n_3 + n_4,  \quad {\rm CH-stretch},   \\
  {\rm v}_2 + {\rm v}_5   &=& n_5 + n_6 + n_7, \quad {\rm CH}_3{\rm-bend}, \\
   {\rm v}_6   &=& n_8 + n_9, \quad {\rm CH}_3{\rm-rock}.
\end{eqnarray*}
Due to the complicated three-step contraction scheme used to construct the symmetrized rovibrational basis set~\citep{Symmetry:2017}, the connection with the primitive basis functions and the assignment based on the largest contribution from them is not straightforward. Therefore, especially in the case of very small values of $|C_i|^2$, the assignment should be considered as indicative.

\begin{table*}
\caption{Extract from the \texttt{.states} file of the OYT-35 line list.}
\vspace*{-3mm}
\begin{threeparttable}
\label{tab:states}
\centering
%\footnotesize
\tabcolsep=5pt
%\resizebox{\linewidth}{!}{
\begin{tabular}{lrcccccccccccccccccc}
\hline\hline\\[-3mm]
        $n$  &  \multicolumn{1}{c}{$\tilde{E}$}   &  $g_{\rm tot}$  &  $J$ & $\Gamma_{\rm tot}$ & $n_1$ & $n_2$ & $n_3$ & $n_4$ & $n_5$ & $n_6$ & $n_7$ & $n_8$ & $n_9$ & $\Gamma_{\rm vib}$ & $J$ & $K$ & $\tau_{\rm rot}$ & $\Gamma_{\rm rot}$ & $|C_i|^{2}$ \\
\hline\\[-3mm]
 1&     0.000000&  16&  0&  1&  0&  0&  0&  0&  0&  0&  0&  0&  0&  1&  0&  0&  0&  1&  0.92\\
 2&   732.842200&  16&  0&  1&  1&  0&  0&  0&  0&  0&  0&  0&  0&  1&  0&  0&  0&  1&  0.89\\
 3&  1354.881100&  16&  0&  1&  0&  0&  0&  0&  0&  1&  0&  0&  0&  1&  0&  0&  0&  1&  0.29\\
 4&  1456.762600&  16&  0&  1&  2&  0&  0&  0&  0&  0&  0&  0&  0&  1&  0&  0&  0&  1&  0.85\\
 5&  2029.375300&  16&  0&  1&  0&  0&  0&  0&  0&  0&  2&  0&  0&  1&  0&  0&  0&  1&  0.13\\
 6&  2080.535700&  16&  0&  1&  1&  0&  0&  0&  0&  1&  0&  0&  0&  1&  0&  0&  0&  1&  0.26\\
 7&  2171.887500&  16&  0&  1&  3&  0&  0&  0&  0&  0&  0&  0&  0&  1&  0&  0&  0&  1&  0.80\\
 8&  2464.902500&  16&  0&  1&  0&  0&  0&  0&  0&  0&  1&  0&  1&  1&  0&  0&  0&  1&  0.22\\
 9&  2694.257786&  16&  0&  1&  0&  0&  0&  0&  1&  1&  0&  0&  0&  1&  0&  0&  0&  1&  0.18\\
10&  2751.022753&  16&  0&  1&  1&  0&  0&  0&  0&  0&  2&  0&  0&  1&  0&  0&  0&  1&  0.12\\
\hline\hline
\end{tabular}
\begin{tablenotes}
\item $n$: State counting number;
\item $\tilde{E}$: Term value (in cm$^{-1}$);
\item $g_{\rm tot}$: Total degeneracy;
\item $J$: Rotational quantum number;
\item $\Gamma_{\rm tot}$: Total symmetry in $\bm{C}_{\mathrm{3v}}\mathrm{(M)}$ (1 is $A_1$, 2 is $A_2$, 3 is $E$);
\item $n_1$\,--\,$n_9$: \textsc{trove} vibrational quantum numbers;
\item $\Gamma_{\rm vib}$: Symmetry of the vibrational contribution in $\bm{C}_{\mathrm{3v}}\mathrm{(M)}$;
\item $J$: Rotational quantum number (same as column 4);
\item $K$: Rotational quantum number, projection of $J$ onto molecule-fixed $z$ axis;
\item $\tau_{\rm rot}$: Rotational parity (0 or 1);
\item $\Gamma_{\rm rot}$: Symmetry of the rotational contribution in $\bm{C}_{\mathrm{3v}}\mathrm{(M)}$;
\item $|C_i^{2}|$: Largest coefficient used in the assignment.
\end{tablenotes}
\end{threeparttable}
\end{table*}

\begin{table}
\centering
\tabcolsep=15pt
\caption{Extract from the \texttt{.trans} file for the $0$\,--\,$100\,$cm$^{-1}$ window of the OYT-35 line list.}
\label{tab:trans}
\begin{tabular}{rrr}
\hline\hline\\[-3mm]
       \multicolumn{1}{c}{$f$}  &  \multicolumn{1}{c}{$i$} & \multicolumn{1}{c}{$A_{if}$}\\
\hline\\[-3mm]
     9719527 & 9719368 & 1.1303e-11 \\
     9719740 & 9719588 & 2.2242e-17 \\
     9720687 & 9910074 & 2.1978e-16 \\
     9722798 & 9532161 & 1.8676e-21 \\
       97590 &   70456 & 2.4642e-07 \\
       97911 &  129246 & 2.7454e-15 \\
       98086 &  129430 & 2.1372e-11 \\
       98104 &  129459 & 5.3801e-12 \\
      981067 & 1101382 & 1.3658e-14 \\
      981676 &  913317 & 4.0655e-10 \\
\hline\hline\\[-2mm]
\end{tabular}

\noindent
\footnotesize{
$f$: Upper  state ID; $i$:  Lower state ID; \\
$A_{if}$:  Einstein-$A$ coefficient (in s$^{-1}$).}
\end{table}

In total, over 334 billion transitions have been computed and this is one of the largest data sets produced by the ExoMol project to date. Although essential for the correct modelling of opacity at elevated temperatures~\citep{14YuTeBa.CH4}, handling such a huge number of lines is cumbersome and a more practical solution is to represent the OYT line lists in a more compact form such as the super-lines format~\citep{TheoReTS:2016,17YuAmTe.CH4}, which is implemented in the \textsc{ExoCross} program~\citep{ExoCross:2018}. To this end, we have generated a set of temperature-dependent super-lines on a non-uniform grid with a constant resolving power of $R= \tilde{\nu}/\Delta \tilde\nu = 1\,000\,000$~\citep{17YuAmTe.CH4} for the $10$\,--\,$6400\,$cm$^{-1}$ region, resulting in 6,461,472 grid points. Super-lines are histograms of the absorption intensities (cm$/$molecule) distributed over this range. Each super-line $k$ is represented by the line position  $\tilde\nu_{k}$ centred at the $k$th bin and the intensity $I_k(T)$  constructed from a sum of intensities $I_{fi}(T)$ of all transitions $f\gets i$ within this bin. As part of the OYT line list package, super-lines have been produced for $T=300$, 400, \ldots, 1100, 1200\,K. Each super-line [$\tilde\nu_{k},I_{fi}(T)$] can be broadened in the standard way to produce pressure-dependent cross-sections with the advantage of a much smaller number of transitions, and a significantly reduced (by 2\,--\,3 orders-of-magnitude) computational cost.

Over 166 billion (166\,279\,320\,228) transitions between 10.2 million (10\,176\,406) energy levels are contained in the OYT-35 line list, while the OYT-37 line list has 168 billion transitions (168\,039\,551\,516) involving 10.2 million (10\,187\,780) states. Figure~2 displays the distribution of lines and energies of the OYT-35 line list as a function of $J$. The largest number of transitions occurs around $J=32$\,--\,$33$ before slowly and smoothly decreasing. For our computational setup, e.g. a lower state energy threshold of $h c \cdot 8000\,$cm$^{-1}$, wavenumber range of $6400\,$cm$^{-1}$, pruned rovibrational basis set, etc., it is apparent that we have not calculated all possible transitions, which would require calculations up to approximately $J=100$. Regarding the number of energy levels the decline after $J=42$ is a consequence of the upper state energy threshold of $h c \cdot 14$\,400\,cm$^{-1}$.

\begin{figure}
\centering
\label{fig:lines_levels}
\includegraphics[width=0.47\textwidth]{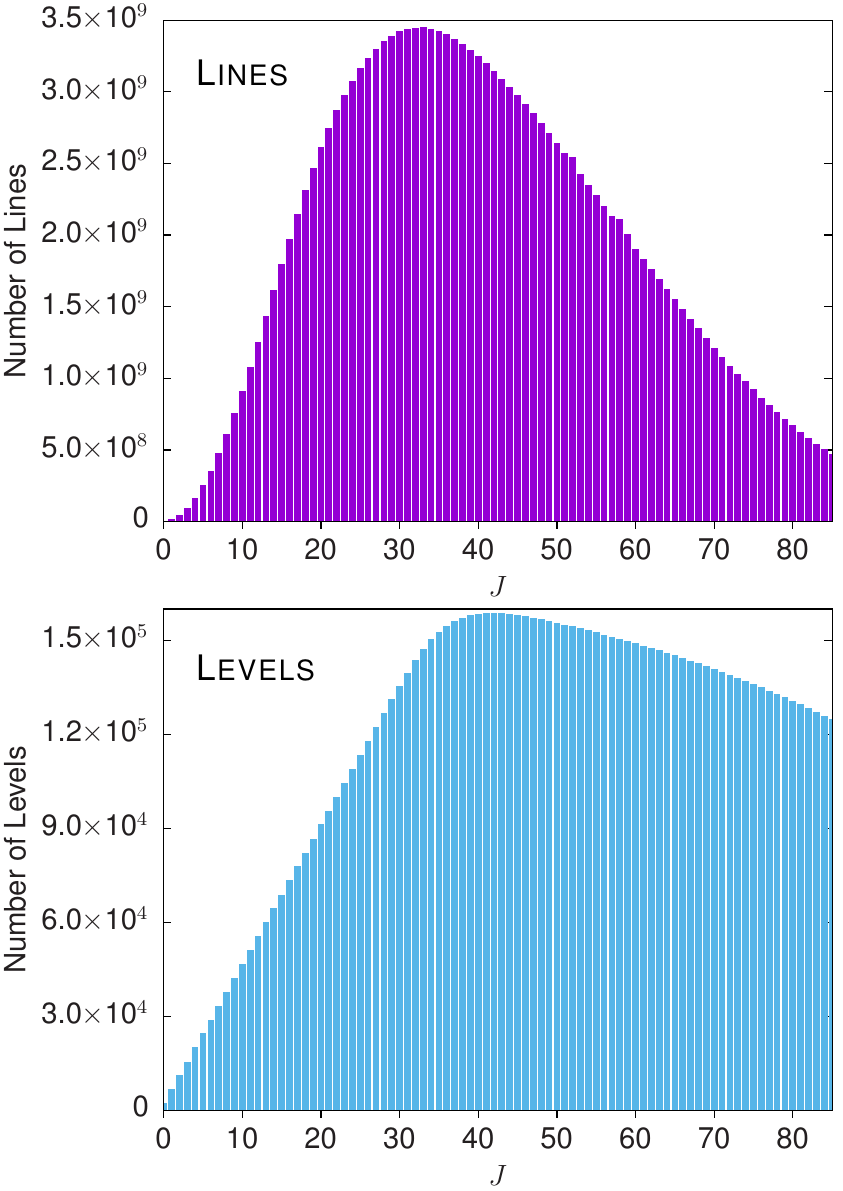}
\caption{The total number of lines and energy levels in the OYT-35 line list for each value of the rotational quantum number $J$. Note that in the upper panel, a single $J$ value counts transitions between $J\leftrightarrow J\!-\!1$, and $J\leftrightarrow J$.}
\end{figure}

\subsection{Partition function of methyl chloride}
\label{sec:pfn}

 The temperature-dependent partition function $Q(T)$ is required for intensity simulations and is defined as,
\begin{equation}
\label{eq:pfn}
Q(T)=\sum_{i} g_i \exp\left(\frac{E_i}{kT}\right) ,
\end{equation}
where $g_i=g_{\rm ns}(2J_i+1)$ is the degeneracy of a state $i$ with energy $E_i$ and rotational quantum number $J_i$. Figure~3  plots the convergence of $Q(T)$ for CH$_3{}^{35}$Cl as a function of $J$ for select temperatures. This was done by summing over all calculated rovibrational energy levels of the OYT-35 line list. Although it is not shown, the same behaviour is exhibited for CH$_3{}^{37}$Cl. The partition function is converged to around $0.2\%$ at $T=1200\,$K. Our computed room temperature partition functions for CH$_3{}^{35}$Cl and CH$_3{}^{37}$Cl are $Q(296\,{\rm K})=57\,973.557$ and $Q(296\,{\rm K})=58\,931.092$, respectively, which are very close to the values from the HITRAN database~\citep{HITRAN2016}. The full partition function for both isotopologues has been computed on a $1\,$K grid between $70$\,--\,$1400\,$K and is given as supplementary material.

\begin{figure}
\centering
\label{fig:pfn}
\includegraphics[width=0.47\textwidth]{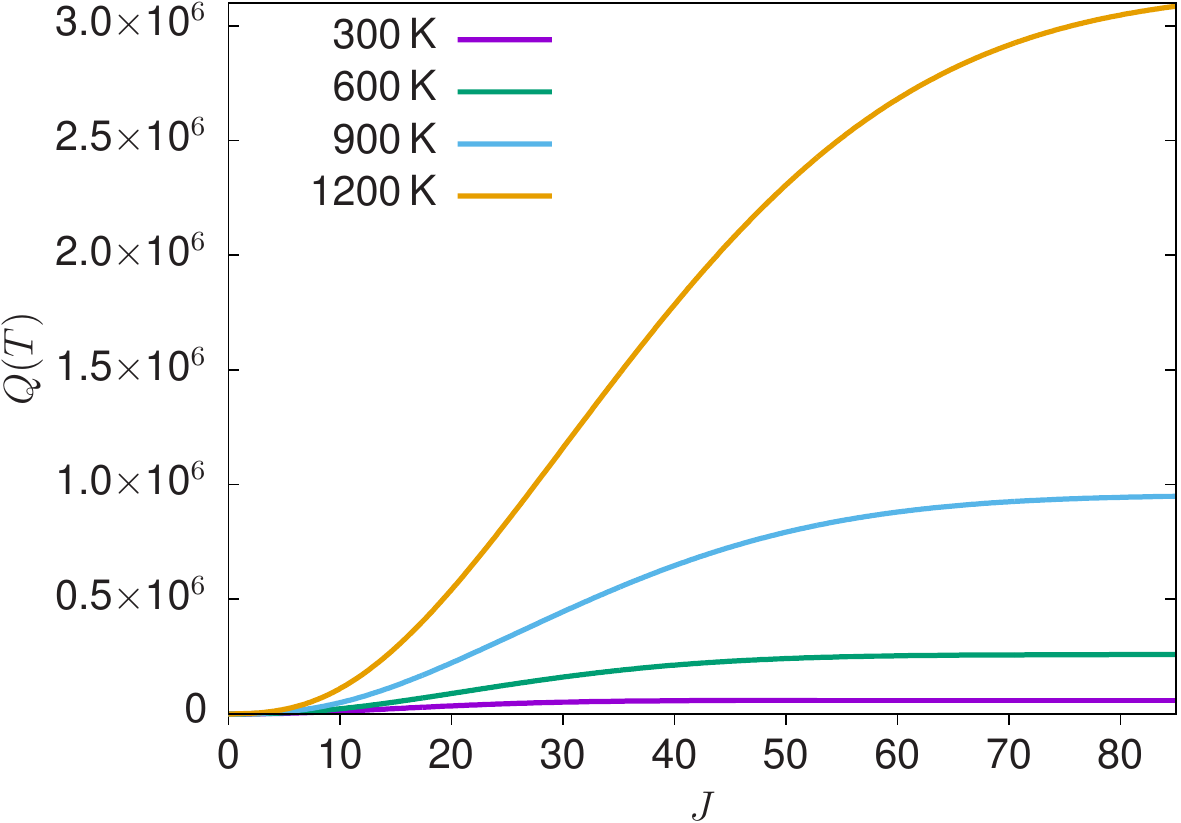}
\caption{Convergence of the partition function $Q(T)$ of CH$_3{}^{35}$Cl with respect to the rotational quantum number $J$ for select temperatures.}
\end{figure}

 Since the OYT line lists have been calculated using a lower state energy threshold of $h c \cdot 8000\,$cm$^{-1}$, it is instructive to look at the reduced partition function $Q_{\rm limit}$, which only includes energy levels up to $h c \cdot 8000\,$cm$^{-1}$ in the summation of Eq.~\eqref{eq:pfn}. For CH$_3{}^{35}$Cl, we have plotted the ratio $Q_{\rm limit}/Q$ with respect to temperature in Fig.~4. This measure allows the completeness of the OYT line lists to be evaluated. At $T=1200\,$K, the ratio $Q_{\rm limit}/Q=0.96$ and this is recommended as a soft temperature limit to the OYT line lists. Above this temperature there will be a progressive loss of opacity when using the OYT line lists but the missing contribution can be estimated through the ratio $Q_{\rm limit}/Q$~\citep{Neale:1996}.

\begin{figure}
\centering
\label{fig:pfn_lim}
\includegraphics[width=0.47\textwidth]{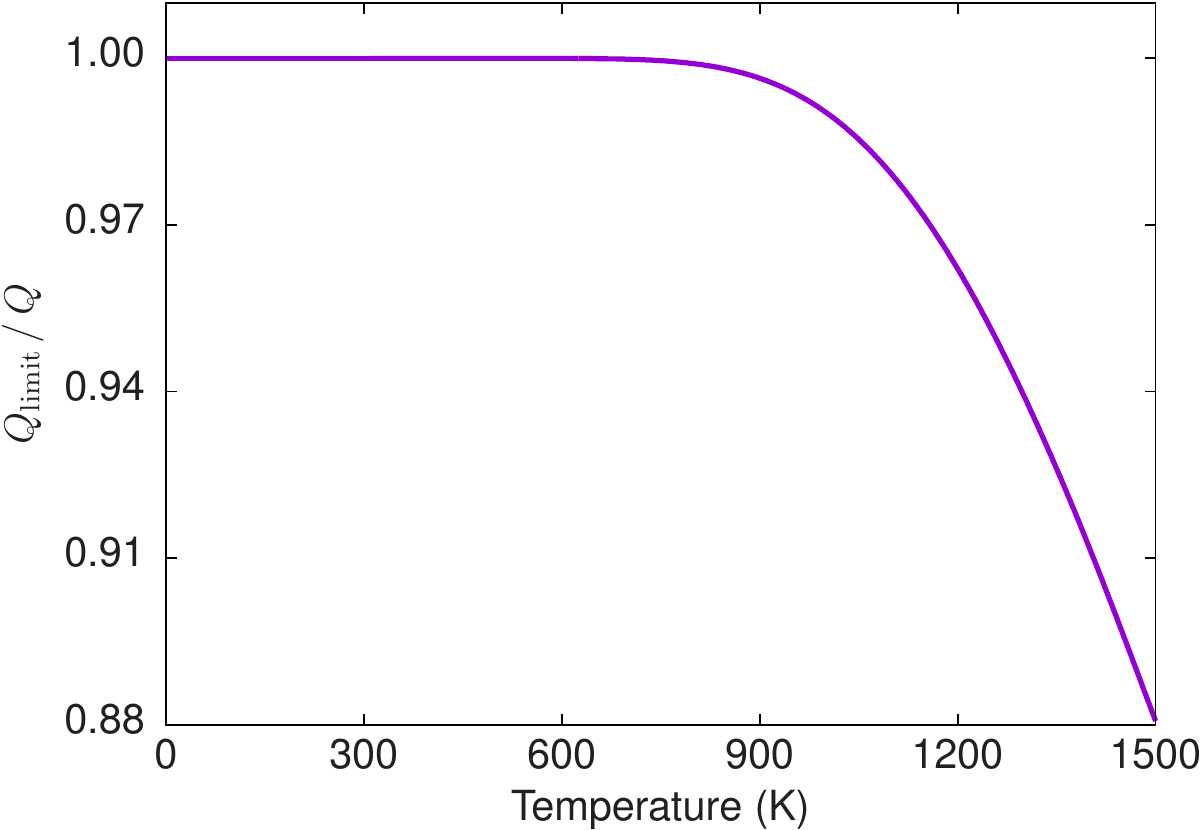}
\caption{The ratio $Q_{\rm limit}/Q$ as a function of temperature; this provides a measure of completeness for the OYT line lists.}
\end{figure}

\subsection{Comparisons with the OYT line lists}
\label{sec:linelist}

 In Fig.~5, integrated absorption cross-sections at a resolution of $1\,$cm$^{-1}$ using a Gaussian profile with a half width at half maximum (hwhm) of $1\,$cm$^{-1}$ have been simulated to illustrate the temperature dependence of the OYT-35 line list. As expected, weak intensities can increase several orders-of-magnitude in strength with rising temperature. This smoothing of the spectrum happens because of the increased population of vibrationally excited states, which causes the rotational band envelope to broaden. Although it is not shown, the OYT-37 line list exhibits identical behaviour.

\begin{figure}
\centering
\label{fig:temp_ch3cl35}
\includegraphics[width=0.47\textwidth]{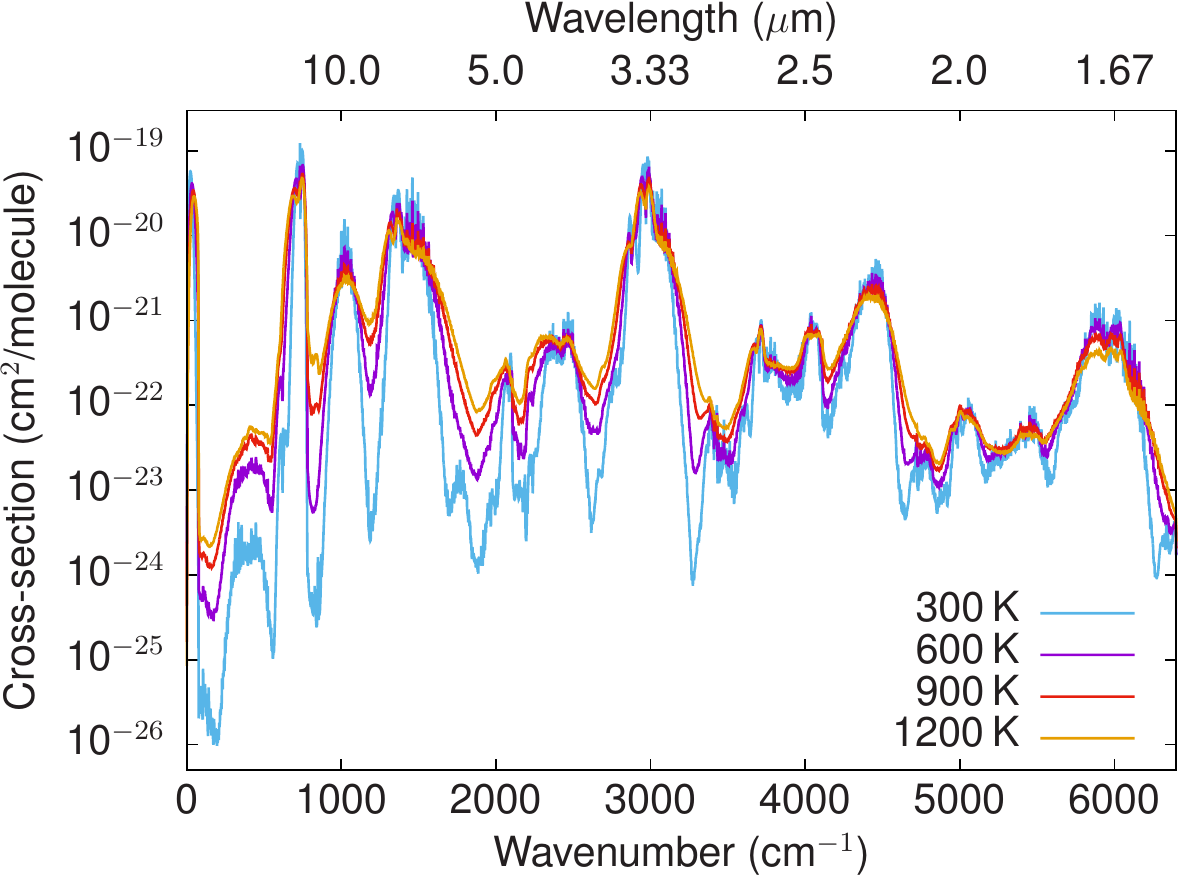}
\caption{The OYT-35 line list temperature dependence: the spectrum becomes increasingly flat as the temperature is raised.}
\end{figure}

 An initial benchmark of the OYT line lists is shown in Fig.~6 and Fig.~7, where we have generated room temperature ($T=296\,$K) absolute line intensities and compared against all lines from the latest HITRAN database~\citep{HITRAN2016}. The OYT intensities have been scaled to natural abundance (0.748\,937 for CH$_3{}^{35}$Cl and 0.239\,491 for CH$_3{}^{37}$Cl), and because the rotational band of CH$_3$Cl is hyperfine resolved in HITRAN, we have `unresolved' the experimental lines to compare with our calculations. As noted previously~\citep{16OwYuYa.CH3Cl}, the only noticeable band missing from HITRAN for wavenumbers below $3200\,$cm$^{-1}$ appears to be the $2\nu_5$ band around $2880\,$cm$^{-1}$, which is not expected to be important for terrestrial atmospheric sensing. Otherwise the agreement is excellent and there are significant improvements compared to our previous efforts~\citep{15OwYuYa.CH3Cl,16OwYuYa.CH3Cl}, particularly regarding line intensities which have benefited from improved variational calculations with an empirically refined PES.

\begin{figure}
\centering
\label{fig:hitran}
\includegraphics[width=0.47\textwidth]{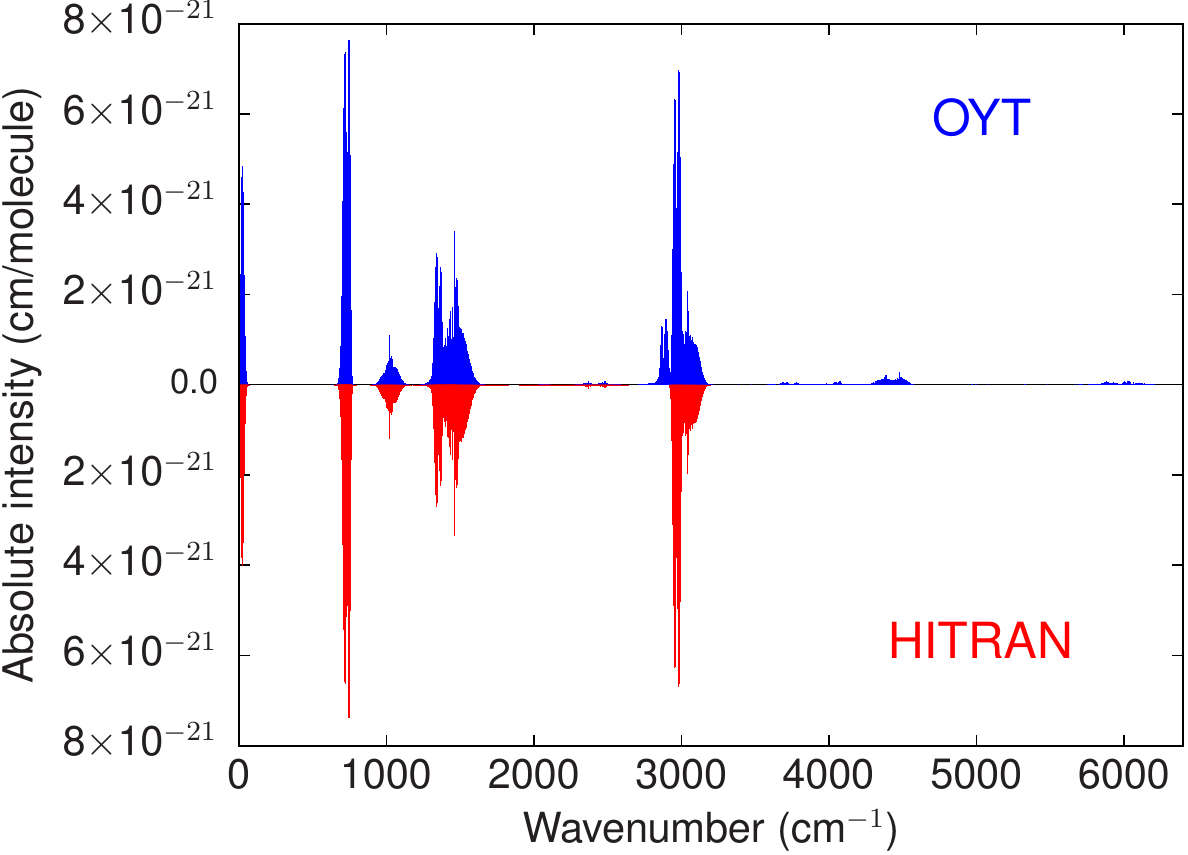}
\caption{Comparison of the OYT line lists against all transitions from HITRAN2016~\citep{HITRAN2016}. The OYT intensities have been scaled to natural abundance. The rotational band from HITRAN has been hyperfine `unresolved' (see text).}
\end{figure}

\begin{figure*}
\centering
\label{fig:hitran_panel}
\includegraphics{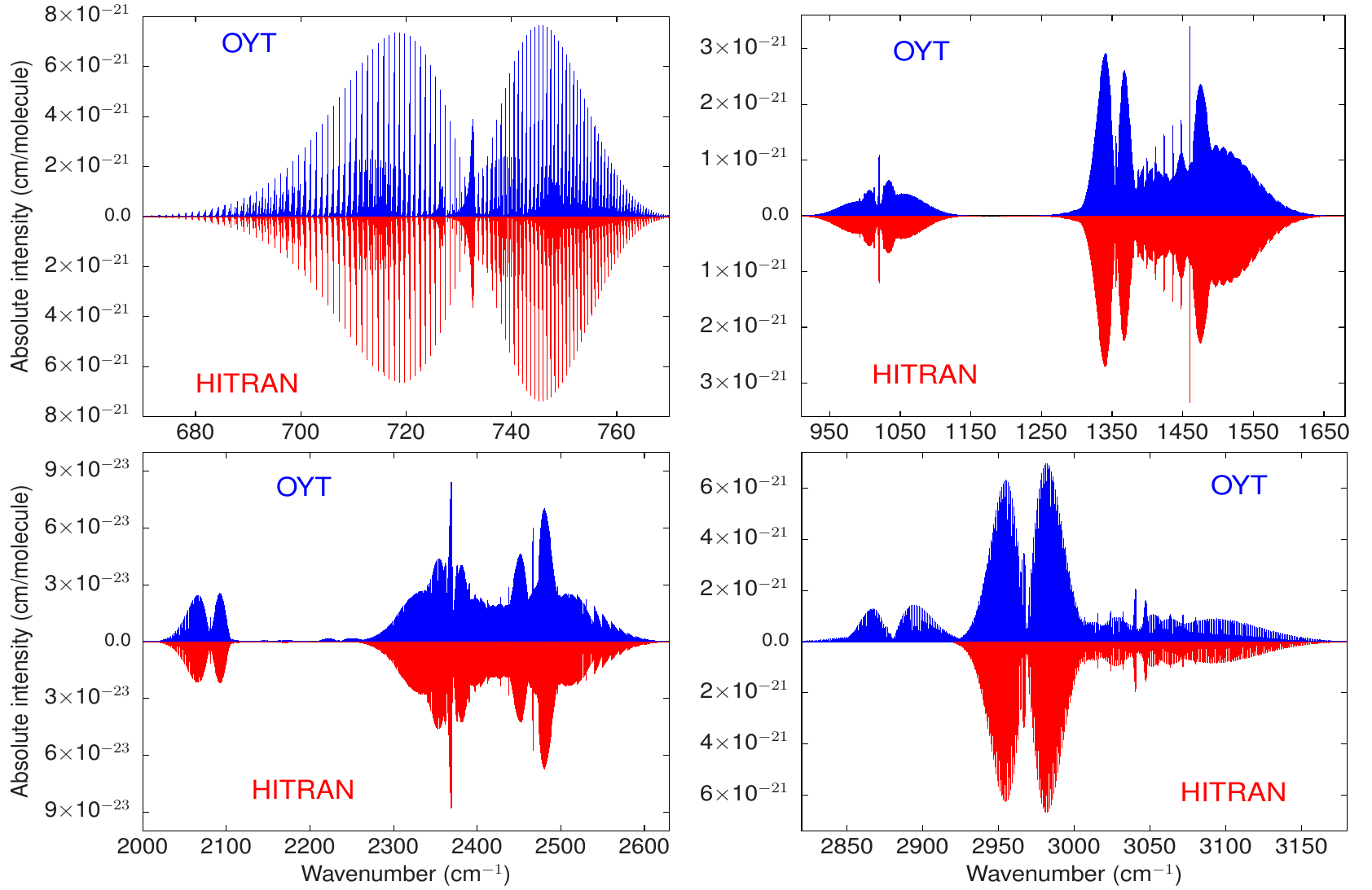}
\caption{The OYT line lists compared with HITRAN2016~\citep{HITRAN2016}. The OYT intensities have been scaled to natural abundance.}
\end{figure*}

 Finally, in Fig.~8 and Fig.~9 we show comparisons with the newly measured high-temperature IR absorption cross-sections at 500$\,^{\circ}{\rm C}$. The OYT spectra are of natural abundance and were simulated at a resolution of $0.1\,$cm$^{-1}$ using a Voigt profile with a Lorentzian line width $\gamma_{\rm L}=0.3\,$cm$^{-1}$. The agreement is extremely pleasing and all the key CH$_3$Cl spectral features are accounted for in the overview presented in Fig.~8. A closer inspection of the bands around $1400\,$cm$^{-1}$, $3000\,$cm$^{-1}$ and $4400\,$cm$^{-1}$ in Fig.~9 provides further proof of the quality of the OYT line lists and confirms the validity of the computational procedure used to construct them.

\begin{figure*}
\centering
\label{fig:fateev_500C}
\includegraphics[width=0.85\textwidth]{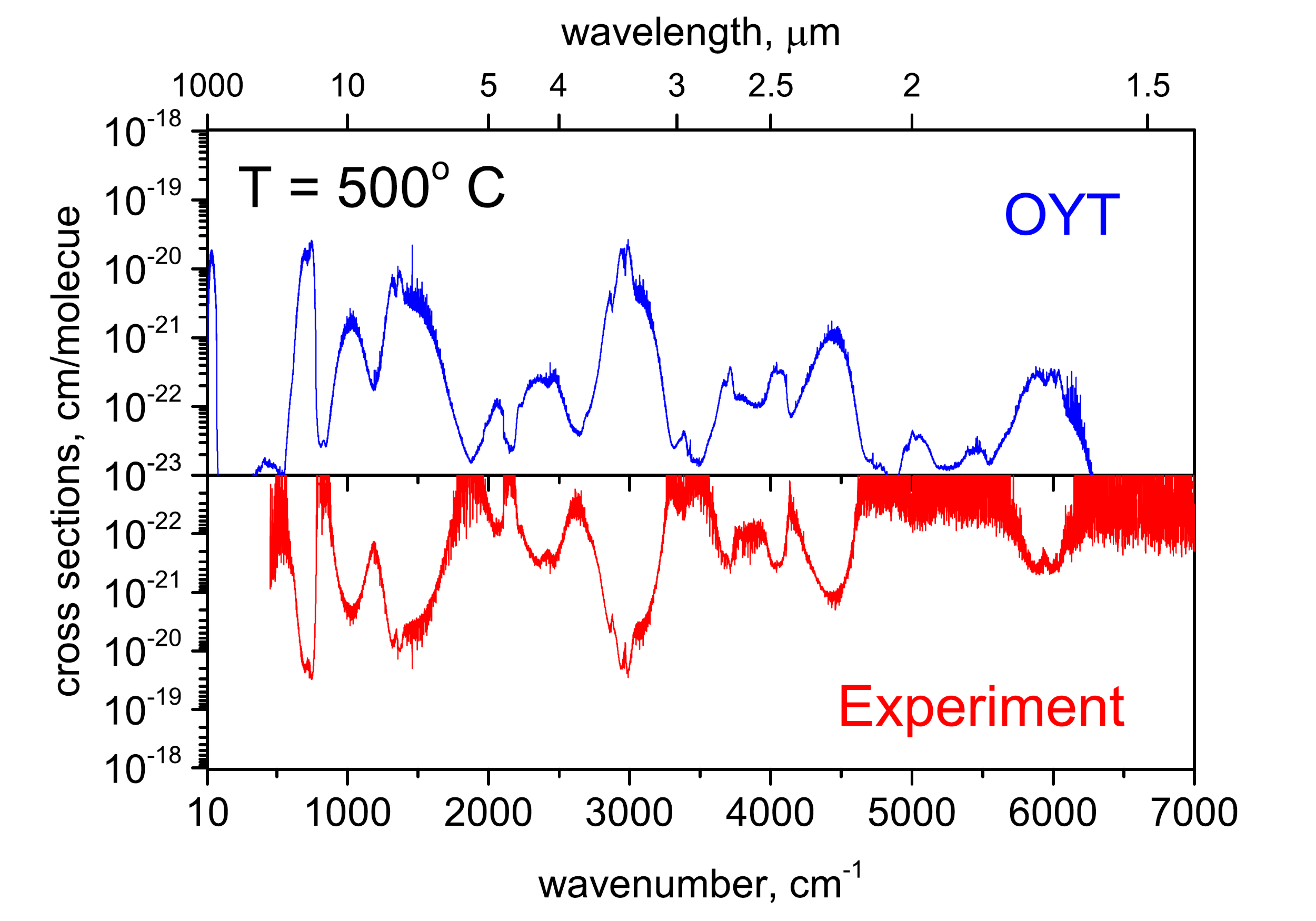}
\caption{OYT absorption cross-sections simulated at 500$\,^{\circ}{\rm C}$ compared with the newly measured IR spectrum. Noise between $4700$\,--\,$7000\,$cm$^{-1}$ is due to the light source and poor MCT detector sensitivity in that region.}
\end{figure*}

\begin{figure}
\centering
\label{fig:fateev_panel}
\includegraphics{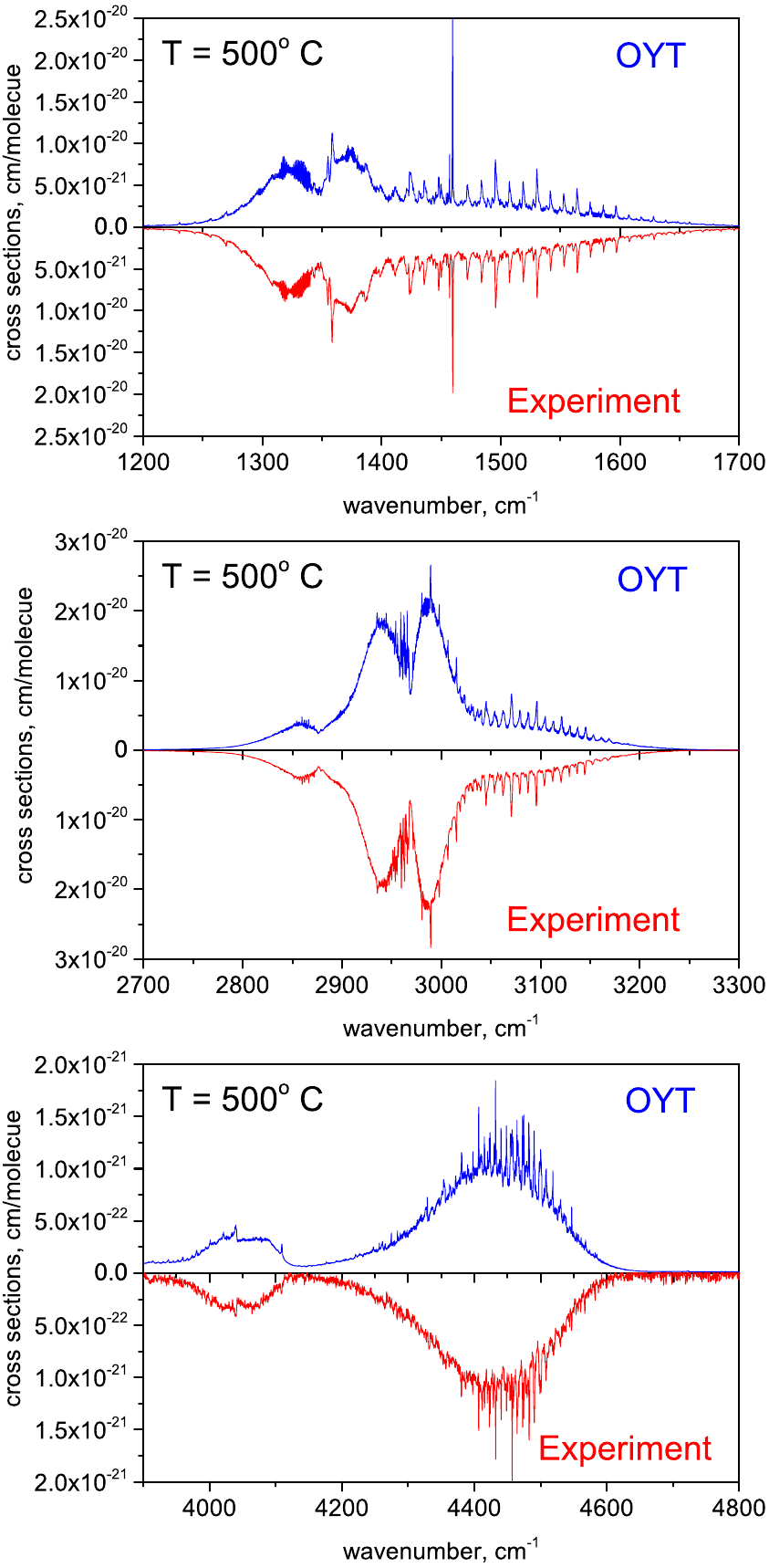}
\caption{OYT absorption cross-sections simulated at 500$\,^{\circ}{\rm C}$ compared with the newly measured IR spectrum.}
\end{figure}

\section{Conclusion}
\label{sec:conc}

 Comprehensive rotation-vibration line lists for the two main isotopologues of methyl chloride, $^{12}$CH$_3{}^{35}$Cl and $^{12}$CH$_3{}^{37}$Cl, have been presented. The OYT-35 and OYT-37 line lists include transitions up to $J=85$ in the $0$\,--\,$6400\,$cm$^{-1}$ range. They are suitable for temperatures up to $T=1200\,$K. Applications above this temperature will lead to the loss of opacity and incorrect band shapes. Comparisons with newly measured high-temperature IR absorption cross-sections confirmed the accuracy and quality of the OYT line lists at elevated temperatures. The line lists are available from the ExoMol database at \url{www.exomol.com} or the CDS database at \url{http://cdsarc.u-strasbg.fr}. 
 
 Possible extensions of the OYT line lists would be an increased lower state energy threshold and frequency range, and the treatment of higher rotational excitation. These issues are relatively straightforward to address, despite being computationally challenging, but will only be done if there is a demand for such work. A complete set of normal mode quantum numbers  $\mathrm{v}_k$ for the OYT line lists would also be useful since these are routinely encountered in high-resolution spectroscopic applications and could be readily incorporated by updating the \texttt{.states} file. Any further updates will be made available on the ExoMol website.

 The completeness and accuracy of the OYT line lists should be adequate for modelling the absorption of methyl chloride in exoplanetary atmospheres. In principle, assuming the abundance of CH$_3$Cl is large enough for detection, transit spectroscopic observations combined with a proper atmospheric and radiative transfer model will be capable of this. For high-resolution detection techniques such as high-dispersion spectroscopy~\citep{Snellen:2014}, the OYT line positions may not be accurate enough. However, hybrid line lists, for example as recently reported for H$_3{}^+$~\citep{Mizus:2017}, can overcome this issue by replacing the computed energy levels with experimentally derived ones, usually obtained with the measured active rotational-vibrational energy levels (MARVEL) procedure~\citep{MARVEL:2007,MARVEL:2012}. Given the amount of experimental spectroscopic data available for CH$_3$Cl a hybrid line list like this could be constructed if necessary.

\section*{Acknowledgments}

This work was part of ERC Advanced Investigator Project 267219. We acknowledge support from the COST Action CM1405 MOLIM, the UK Science and Technology Research Council (STFC) No. ST/M001334/1  and the Max Planck Computing and Data Facility (MPCDF). A part of the calculations were performed using DARWIN, high performance computing facilities provided by DiRAC for particle physics, astrophysics and cosmology and supported by  BIS National E-infrastructure capital grant ST/J005673/1 and STFC grants ST/H008586/1, ST/K00333X/1. A.O.\ and A.Y.\ acknowledge support from the \emph{Deutsche Forschungsgemeinschaft} (DFG) through the excellence cluster ``The Hamburg Center for Ultrafast Imaging -- Structure, Dynamics and Control of Matter at the Atomic Scale'' (CUI, EXC1074). A.F.\ acknowledges support from Energinet.dk (project 2013-1-12027 ``High-resolution spectroscopy of Cl-compounds in gasification'').

\bibliographystyle{mn2e}
%\bibliography{exomol_ch3cl_01}

\section*{Supporting Information}
Supplementary data are available at MNRAS online.

\label{lastpage}

\end{document}